\documentstyle[12pt,graphicx]{article}

\begin{document}
\thispagestyle{empty}
\hfill{\small KL--TH 97/5}\\[10mm]
\begin{center}
{\Large\bf Calculation of Spin Tunneling Effects in the Presence of an Applied
Magnetic Field}\\[16mm]
{\large J.--Q. Liang, H.J.W. M\"uller--Kirsten,  A.V. Shurgaia and F.Zimmerschied}\\[10mm]
{\it Department of Physics, University of Kaiserslautern, D--67653
Kaiserslautern, Germany,} \\[4cm]
{\bf Abstract}
\end{center}

The tunneling splitting of the energy levels of a ferromagnetic particle in the presence
of an applied magnetic field -- previously derived only for the ground state with
the path integral method -- is obtained in a simple way from Schr\"odinger theory.  The
origin of the factors entering the result is clearly understood, in particular the 
effect of the asymmetry of the barriers of the potential. The method should appeal 
particularly to experimentalists searching for evidence of macroscopic spin tunneling.

\vspace{3cm}
\begin{center}
\end{center}
\newpage       

In spite of early speculation\cite{1} that the one loop approximation of a path integral
calculation in quantum mechanics would not yield exactly the same result as a WKB
semiclassical calculation, it is now generally agreed -- and has indeed been verified
by application to
important examples -- that the two approaches yield exactly the same result in the
approximation of correct linear matching of WKB solutions across
turning points.  Thus with this proviso the latter method was shown \cite{2} to
lead to the same level shifts in the cases of the double--well
and periodic potentials as the path integral method with expansion
around an instanton.  The effect of the nonlinearity which gives rise to
instanton or other classical solitonic configurations arises in
Schr\"odinger quantum mechanics through the boundary conditions imposed on the
perturbatively calculated wave functions.  It is these boundary conditions therefore
which generate the nonperturbative effects characteristic also of tunneling.

Once the Hamiltonian of a theory with tunneling effects has been established, it is still a long way
to obtain the level shift or decay rate with the path integral method.
Such calculations are very instructive, particularly as models for more
complicated field theoretical contexts, but can lack transparency
in other contexts where a confrontation with experiments is anticipated.
Thus recently long and complicated instanton path integral
calculations \cite{3,4,5,6,7,8,9,10,11} were performed in order to derive the tunneling
splitting of the ground state of a magnetic particle in a crystal field
with or without an applied magnetic field.  This is a case which is
of considerable experimental interest since the tunneling rate should be directly experimentally 
measurable in resonance experiments\cite{11}.  It seems it is not so well known
that the calculation of such tunneling effects can also be performed in the
context of Schr\"odinger theory with appropriately chosen (in this case periodic)
boundary conditions, and in fact much more directly so by comparison
with established cases as demonstratred recently in the case with no applied
magnetic field\cite{12}. On the basis of our experience with both
types of methods (for path integral calculations see also refs.\cite{13,14,15}), we
expect that any calculation which is manageable in the path integral method
can also be dealt with in Schr\"odinger theory and vice versa.  In fact, the
latter even has certain advantages in yielding with equivalent ease
(or complexity) the corresponding results for excited states whereas in
the path integral method this would necessitate the consideration of periodic or nonvacuum
instantons\cite{14,15} which implies a further degree
of considerable calculational complication.

Quantum spin tunneling (or more generally tunneling for internal degrees of
freedom) has only recently been realised to be a fascinating quantum mechanical
phenomenon. In the following we demonstrate that the results obtained
in refs.\cite {3,4} and \cite{11}
for spin tunneling between classically degenerate states in the presence
of an applied magnetic field can essentially be obtained in a very simple
way by comparison of the Schr\"odinger equation with that of a periodic 
potential for which the level splitting is well known, i.e. the Mathieu
equation\cite{16}.  Apart from supplying a check of the path integral
result, this type of derivation should appeal particularly to
experimentalists.  Moreover it has the enormous advantage of
yielding automatically the extension to excited states which -- as remarked  earlier --
is much more difficult to achieve with the path integral method.  The calculation also
gives a clearer picture of which parameters have to be large and which small
for the results to have validity in the sense of asymptotic expansions. 

In the following we concentrate on the case of a magnetic particle (e.g.
an ion) embedded in the field of a crystal and a weak applied magnetic field $h = g\mu_B.{\cal B}
$ perpendicular to the easy axis $X$. 
As explained in refs.\cite{3,4} for a specific model (among others)
the appropriate Hamiltonian is
\begin{equation}
{\hat H} = -A{\hat S}^2_x + B{\hat S}^2_z - h{\hat S}_y
\label{1}
\end{equation}
where $A$ and $B$ are positive anisotropy constants.
One should note that two anisotropy axes are needed to produce tunneling
when $h = 0$.
After transforming the
spin operators ${\hat S}_i$ to canonical operators ${\hat p}, {\hat \phi}$ \cite{17}
the relevant semiclassically approximated Hamiltonian can be written
\begin{equation}
{\hat H} = \frac{{\hat p}^2}{2m({\hat {\phi}})} + V({\hat {\phi}})
\label{2}
\end{equation}
where $[{\hat p}, {\hat \phi}] = i$ and 
\begin{equation}
m(\phi) =\left[2(A\cos^2\phi + B) + {\frac{h\sin\phi}{s+\frac{1}{2}}}\right]^{-1}
\label{3}
\end{equation}
and
\begin{equation}
V(\phi) = -\left[As(s+1)\cos^2\phi + h(s+\frac{1}{2})\sin\phi\right]
\label{4}
\end{equation}
where $s$ is the spin of the particle which is assumed to be not too small.
The classical phase space is the unit sphere.  The potential (which is two--fold degenerate
for $h=0$ in the domain $0\leq \phi \leq\pi$ with easy axis X, i.e. the degenerate
minima at $0$ and $\pi$) can be seen to result from the Hamiltonian ${\hat H}$
obtained by replacing ${\hat S}_i$ by $s{\bf e}_i(\theta, \phi)$, ${\bf e}_i$ a unit
vector expressed in polar coordinates, and replacing $\theta$ by its classical
ground state value $\frac{\pi}{2}$.  The derivation of the field dependence of the mass
(which does not concern us here) is described in refs.\cite{3,4}. 
In the following we set
\begin{equation}
b = \frac{B}{A},\;\;\; a = \frac{h}{2A\left(s+\frac12\right)}
\label{5}
\end{equation}
and consider the experimentally interesting case of $b>1$.  Thus the mass $m(\phi)$
is to a first approximation given by $\frac{1}{2B}$.  We demonstrated earlier\cite{12}
that corrections to this are small and can be taken into account
in terms of elliptic functions (a modification which could, in principle, also be considered
here but would overload this note with algebraic
technicalities). We ignore these corrections in the first place. 

The quantum states of the system (i.e. the degenerate states separated by
infinitely high barriers) are determined by the oscillator
approximation of the system around minimum positions $\phi_0$ of $V$ given by
(with $s(s+1)\approx \left(s+\frac{1}{2}\right)^2$) 
\begin{equation}
\sin\phi_0 = a, \;\;\; \cos\phi_0 =\pm\sqrt{1-a^2}
\label{6}
\end{equation}
and so
\begin{equation}
\phi_{0,2n} =  2n\pi +\arcsin a, \;\;\; \phi_{0,2n+1} = (2n+1)\pi-\arcsin a, \;\;\;n = 0,1,2,...
\label{7}
\end{equation}
with
$$\cos\phi_{0,2n} = +\sqrt{1-a^2}, \;\;\; \cos\phi_{0,2n+1} =  -\sqrt{1-a^2}
$$
at which
\begin{equation}
V^{\prime\prime}(\phi_0) = 2As(s+1)[1-a^2] \equiv 8h^2_m.B
\label{8}
\end{equation}
In the problem at hand $0\leq \phi \leq 2\pi$ and there are only two
minima as in the case of the well known double well potential.
The Schr\"odinger equation ${\hat H}\Psi = E\Psi$ defined by ${\hat H}$
is then approximately
\begin{equation}
\Psi^{\prime\prime}(\phi) + \left[\frac{E-V(\phi_0)}{B} - 4h^2_m(\phi - \phi_0)^2\right]\Psi\simeq 0
\label{9}
\end{equation}
This determines immediately the oscillator approximated eigenvalues as
\begin{equation}
E^{(0)}_{2n+1}= -As(s+1) - \frac{h^2}{4A} + (2n+1)\sqrt{AB}
\left(s+\frac12\right)\sqrt{1-a^2}
\label{10}
\end{equation}
in agreement with the ``semiclassical ground state energy'' ($n=0$) given in ref.\cite{3}
(last formula) except that our $B$ (assumed to be much larger than $A$) is there $B+A$.
This perturbation theoretical expression ignores tunneling.

\begin{figure}
\begin{center}
\includegraphics[bb= 3cm 8.9cm 20cm 22cm,clip,scale=0.6]{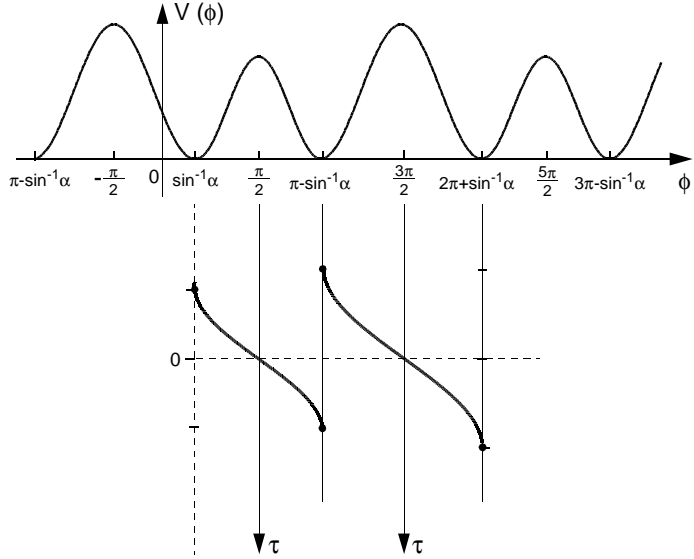}
\end{center}
\caption{The periodic potential with asymmetric twin barriers and the instanton trajectories.}
\vspace{6mm}
\label{fig1}
\end{figure}

A typical aspect of any tunneling formula is the exponential of the euclidean
action of the classical vacuum pseudoparticle (which tunnels through the barrier).  This
factor supplies the classical approximation of the transition amplitude 
equivalent to the wave function approximation given by the WKB exponential in
quantum mechanics and must be such that it vanishes (i.e. the level splitting)
in the limit of infinitely high barriers.  In the present case this implies that
$h^2_m$ has to be large and thus $s$ or $As(s+1)$.  The argument of the 
exponential must therefore contain $s$.  In the path integral results of
refs.\cite{3,4} the factors raised to powers of $2\left(s+\frac12\right)$ 
play this role.  In fact the
factor in eq. (9a) of ref.\cite{3} can be shown to approximate the
WKB factor $exp(-\frac{2\left(s+\frac12\right)}{\sqrt{b}})$ in the 
limit of vanishing magnetic
field $h$ (one may note that $a=1$ determines a critical value of $h$
beyond which the degeneracy of minima for $h\neq 0$ -- cf. eq.(\ref{6}) --
is removed as discussed by Schilling\cite{11}).
These observations suggest that the tunneling effect (i.e. the
lifting of the degeneracy of oscillator levels) is calculable in much the same way
as for periodic potentials, and that in fact the result can be obtained by
identification of appropriate parameters.  We do not enter into an
extensive calculation here along the lines of ref.\cite{16}, and instead
proceed by identification.  One avoids ugly integrals by setting in the 
Schr\"odinger equation $E = E^{(0)}_{2n+1} + \triangle$, where
$\triangle$ is the perturbation theory correction of the eigenvalue.  The
original Schr\"odinger equation then becomes
\begin{equation}
\Psi^{\prime\prime}+\left\{-G^2(\phi) + (2n+1)\sqrt{\frac{A}{B}}
\left(s+\frac12\right)\sqrt{1-a^2} +\frac{\triangle}{B}
\right\}\Psi = 0
\label{11}
\end{equation}
where
\begin{equation}
G^2(\phi) = \frac{\left(s+\frac12\right)^2}{b}(\sin\phi - a)^2
\label{12}
\end{equation}
Setting
$$\Psi = \Xi(\phi)\exp\left\{\pm\int^{\phi}G(\phi)d\phi\right\}$$
we obtain the WKB exponential
\begin{equation}
\exp\left\{\pm\int^{\phi}G(\phi)d\phi\right\} = \exp\left\{\pm 
\frac{s+\frac12}{\sqrt{b}}(\cos\phi + a\phi)\right\}
\label{13}
\end{equation}
(note that without inclusion of the approximate form of $E$ the integral would not have 
been so simple !).  
One can compute $\Xi(\phi)$ etc. as in ref.\cite{16}.  The boundary conditions
require the evaluation of the wave function
above the chosen minimum of the potential
(say that at $\phi_{0,1}$), implying for the barriers to the left and to
the right
\begin{eqnarray}
\int^{\phi_{0,1}}G(\phi)d\phi -\int^{\phi_{0,0}}G(\phi)d\phi =
\frac{2\left(s+\frac12\right)}{\sqrt{b}}\left(\sqrt{1-a^2} + 
a\arcsin{a}-a\frac{\pi}{2}\right)\nonumber\\
\int^{\phi_{0,1}}G(\phi)d\phi - \int^{\phi_{0.2}}G(\phi)d\phi = 
\frac{2\left(s+\frac12\right)}{\sqrt{b}}\left(
\sqrt{1-a^2} + a\arcsin{a} + a\frac{\pi}{2}\right)
 \label{14}
\end{eqnarray}
These expressions are seen to be (cf.ref.\cite{10}) precisely
the values of the action of the instantons travelling through the two differently
sized barriers between $(\phi_{0,1}, \phi_{0,0})$
and $(\phi_{0,1}, \phi_{0,2})$ respectively as shown in Fig.\ \ref{fig1}
(one could say that these
instantons describe clockwise and anticlockwise underbarrier
rotations of the magnetic moment). In order to obtain this result it was
essential to insert the oscillator approximated energy eq.(\ref{10}) into the 
Schr\"odinger equation.  It is exponential factors like those of eq.(\ref{13}) with
the boundary conditions of eq.(\ref{14})
which are typical of
tunneling contributions.  In the present case both of these contribute to the overall
level splitting.  Knowing these factors we can     
write down the level splitting by making the appropriate replacements in the
formula for the level splitting in the case of the Mathieu equation (cf. ref.\cite{16}
,also cited in \cite{12})
and adding these with equal weights so that in the limit $a\rightarrow 0$ we
regain the level splitting of the case without the magnetic field.  The factors multiplying
the exponentials are characteristic of the central well (i.e.
$h^2_m$ of eq.(\ref{8})) and in fact result from the matching of different branches
of the wave function in domains of overlap as can be seen from ref.\cite{16};
classically this factor describes the number of bounces of the particle between the
barriers before it escapes. 
Thus these factors are the same in both cases so that from the level splitting result
of ref.\cite{16} (also cited in \cite{12}) the level splitting of the $n$th excited state
in the present case is obtained by the replacement
$$
e^{-\frac{2\left(s+\frac12\right)}{\sqrt{b}}} 
\rightarrow \frac{1}{2}e^{-{\frac{2s}{\sqrt{b}}}\left(\sqrt{1-a^2}
+a\arcsin{a}\right)}.\left[e^{\frac{a\left(s+\frac12\right)\pi}{\sqrt{b}}} + 
e^{-\frac{a\left(s+\frac12\right)\pi}{\sqrt{b}}}\right]
$$
and hence is found to be
\begin{equation}
\triangle_{2n+1}=\frac{2B\left[\frac{8\left(s+\frac12\right)}{\sqrt b}
\sqrt{1-a^2}\right]^{n+\frac{3}{2}}}{\sqrt{8\pi}n!}\cdot
e^{-\frac{2\left(s+\frac12\right)}{\sqrt b}(\sqrt{1-a^2}+a\arcsin a)}\cdot
\cosh(\frac{a\left(s+\frac12\right)\pi}{\sqrt b})
\label{15}
\end{equation}
In the limit $a\rightarrow 0$ this reduces to the formula obtained
in ref.\cite{12} or to $\triangle E^{inst}_0$ of formula (9a) of ref.\cite{3}.
In particular we obtain for $n=0$
\begin{equation}
\triangle_1 = \frac{16B}{\sqrt{\pi}}\left[\frac{s+\frac12}{\sqrt{b}}
\sqrt{1-a^2}\right]^{\frac{3}{2}}\cdot
e^{-\frac{2\left(s+\frac12\right)}{\sqrt{b}}(\sqrt{1-a^2}+a\arcsin a)}
\cdot\cosh(\frac{a\left(s+\frac12\right)\pi}{\sqrt{b}})
\label{16}
\end{equation}
which is to be compared with the corresponding path integral result of ref.\cite{4}
(there eq.(16)).  In our result the origin of every factor is clearly understood as
explained above - of course, with the assumption $B>A$ and $s>1$. The somewhat
different factors in ref.\cite{4} result from the complicated path integral calculation, taking
into account of the field dependence of the mass and appropriate approximations.

\begin{figure}
\begin{center}
\includegraphics[bb= 2cm 9.1cm 18cm 22cm,clip,scale=0.6]{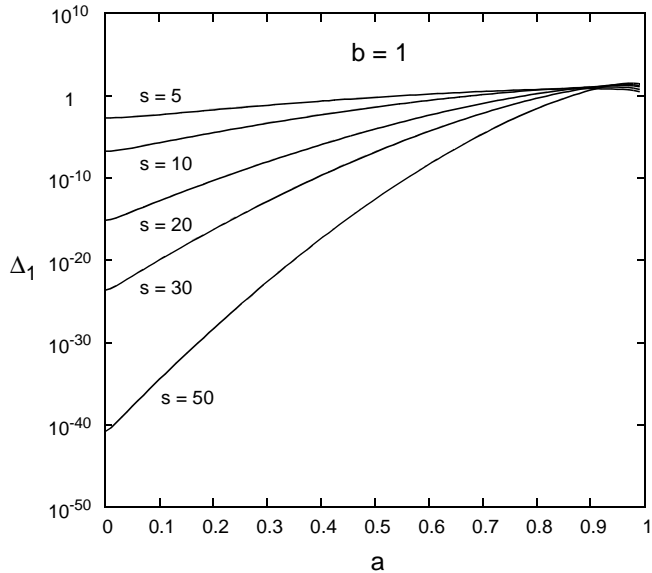}
\end{center}
\caption{Logarithmic plot of $\triangle_1$ (in units of $A$) as a 
function of $a$ for $b=\frac{B}{A}=1$
and $s = 5, 10, 20, 30, 50$ to be compared with Fig. 2 of ref.\cite{4}.} \vspace{6mm}
\label{fig2}
\end{figure}

In Fig.\ \ref{fig2}  we plot our expression $\triangle_1$ for the values given in Fig. 2 of
ref.\cite{4}.  We make the amazing observation
that the plots are practically identical in spite of the fact that
the value of $b$ chosen (i.e. $b = 1$) is really too small and still better
agreement with the exact values will be obtained
for larger values of $b$. 

\begin{figure}
\begin{center}
\includegraphics[bb= 1cm 8.6cm 17.5cm 21.5cm,clip,scale=0.5]{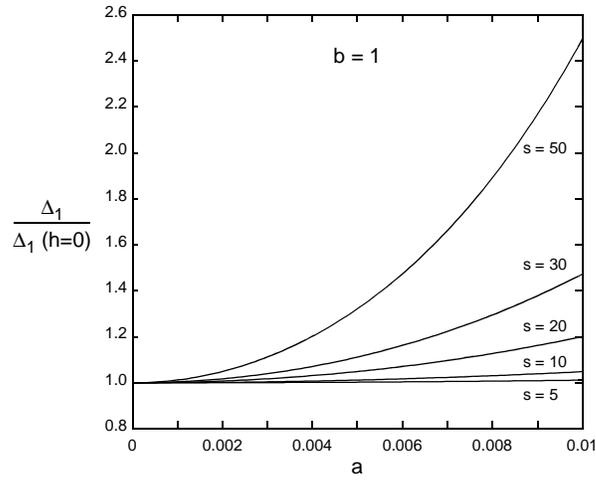}
\end{center}
\caption{The ratio of the level splitting divided by that for $h = 0$ (i.e. 
magnetic field zero) plotted
against $a$. One should note that $a$ is effectively $h$ divided by the large spin $s$. } 
\vspace{6mm}
\label{fig3}
\end{figure}

\begin{figure}
\begin{center}
\includegraphics[bb= 4.5cm 10cm 18cm 23cm,clip,scale=0.5]{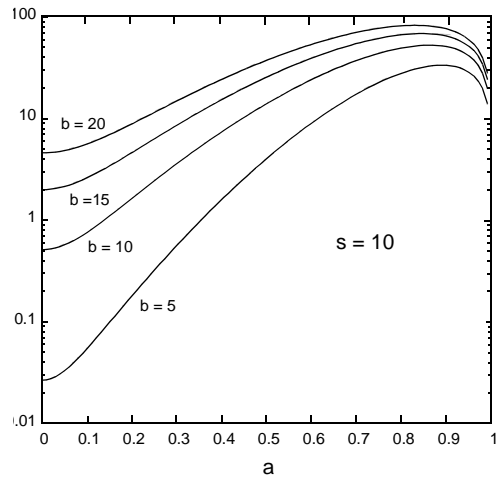}
\end{center}
\caption{The level splitting at fixed value $s = 10$ for $b = 5, 10, 15$ and $20$. 
In our considerations here
$b$ has to be larger than $1$.} \vspace{6mm}
\label{fig4}
\end{figure}

In Figs.\ref{fig3} and \ref{fig4} we display further plots of the level
splitting demonstrating its increase with the magnetic field (as desired for
better observability) and its variation with $b$ for a fixed value of $s$.

In Table \ref{tab1} we display some absolute values of the level splitting as
calculated from our result and compare these with values given in ref.\cite{3}.  

\begin{table}
\begin{center}
\begin{tabular}{|c|c|c|c|}
\hline
$h$ & $\Delta_1$ & $\Delta_1/\Delta E_0$ & $\Delta E_0^{inst}/\Delta E_0$  \\
\hline\hline
0.0 & $2.4488\times 10^{-4}$ & $1.104$ & $1.029$ \\ \hline
0.4 & $3.4375\times 10^{-4}$ & $1.110$ & $1.065$ \\ \hline
0.8 & $7.0830\times 10^{-4}$ & $1.111$ & $1.123$ \\ \hline
1.2 & $1.5803\times 10^{-3}$ & $1.145$ & $1.174$ \\ \hline
\end{tabular}
\end{center}
\caption{Level splitting values calculated from
$\Delta_1$ in units of $A$ compared  with the numerical values $\Delta E_0$ and the
semiclassical results $\Delta E_0^{inst}$ of ref.
\cite{3} for $s=10$, $b=2$ and different values of $h$}
\label{tab1}
\end{table}

The very simple derivation of the nontrivial level splitting of eq.(\ref{16}) given here demonstrates
the calculational superiority of the Schr\"odinger method in quantum mechanics. 
Of course, the path integral method has considerable 
pedagogical value in quantum mechanics as a model for more complicated
applications in field theory.
The derivation given here also shows clearly that the parameters $s$ and $h^2_m$
have to be large for the asymptotic expansions of the solutions and eigenvalues
to make sense.  One can also see that the effect of the field dependence of the mass
-- here neglected -- cannot be large (in agreement with our findings in ref.\cite{12}).
A further considerable advantage of the derivation given here is -- apart from checking the path
integral result -- that it immediately supplies the splitting of higher
degenerate oscillator levels which cannot be obtained with vacuum instanton
methods.  It is interesting to observe that the WKB exponential which corresponds to the classical
or tree approximation is the same also for excited states.  Thus if this is calculated 
for nonvacuum (or periodic) instantons the same expression must be obtained.
We did not specify above whether $s$ is integral or half--integral. The reason
is that for half integer spins the classical twofold degeneracy
remains due to Kramer's degeneracy \cite{8}.
For completeness we mention that the case $B=0$ has been dealt with in ref.\cite{18},
the method there employed being one applicable to the original
discrete spin system.
The method discussed here obviously also shows how the level splitting in the
case of a periodic potential with periodically recurring asymmetric twin barriers
can be obtained. We remark that numerous properties of spin systems have been
considered in \cite{19}. 
Finally our presentation should also appeal to experimentalists in view of
the transparency of the calculational steps.

\vspace{2cm}
  
\centerline{\bf Acknowledgements}

\vspace{0.3cm}
Two of us, J.--Q.L. (permanently at Department of Physics, Shanxi University,
Taiyuan, P.R. China) and A.V.S. (permanently at Department of Mathematics, Georgian
Academy of Sciences, Tbilisi, Georgia) are indebted to the Deutsche Forschungsgemeinschaft
for financial support. The collaboration has also been supported by INTAS--93--1630 EXT.

\end{document}